\documentclass{article}

\usepackage{amssymb,amsfonts,amsmath,stmaryrd}
\usepackage{cite,enumerate,float,indentfirst}
\usepackage{color}

\def\be{\begin{eqnarray}}
\def\ee{\end{eqnarray}}
\def\nn{\nonumber}

\def\p{\partial}

\newcommand{\beq}{\begin{equation}}
\newcommand{\eeq}{\end{equation}}
\newcommand{\beqa}{\begin{eqnarray}}
\newcommand{\eeqa}{\end{eqnarray}}

\definecolor{red}{rgb}{1,0,0}
\definecolor{orange}{rgb}{1,0.5,0}
\definecolor{violet}{rgb}{0.7,0,1}

\newcommand{\ttop}[1]{
  q^{\hat{D}_#1}
}

\voffset= - 1.1in
\hoffset= - 1.0in         

\begin{document}

\title{\vspace{1.5cm}\bf
On Chalykh's approach to eigenfunctions of \\ DIM-induced integrable Hamiltonians
}

\author{
A. Mironov$^{b,c,d,}$\footnote{mironov@lpi.ru,mironov@itep.ru},
A. Morozov$^{a,c,d,}$\footnote{morozov@itep.ru},
A. Popolitov$^{a,c,d,}$\footnote{popolit@gmail.com}
}

\date{ }

\maketitle

\vspace{-6cm}

\begin{center}
  \hfill MIPT/TH-23/24\\
  \hfill FIAN/TD-14/24\\
  \hfill ITEP/TH-29/24\\
  \hfill IITP/TH-24/24
\end{center}

\vspace{4.5cm}

\begin{center}
$^a$ {\small {\it MIPT, Dolgoprudny, 141701, Russia}}\\
$^b$ {\small {\it Lebedev Physics Institute, Moscow 119991, Russia}}\\
$^c$ {\small {\it NRC ``Kurchatov Institute", 123182, Moscow, Russia}}\\
$^d$ {\small {\it Institute for Information Transmission Problems, Moscow 127994, Russia}}
\end{center}

\vspace{.1cm}

\begin{abstract}
Quite some years ago, Oleg Chalykh has built a nice theory from the
observation that the Macdonald polynomial reduces at $t=q^{-m}$ to a sum over permutations of simpler polynomials called Baker-Akhiezer functions, which can be unambiguously constructed from a system of linear difference equations. Moreover, he also proposed a generalization of these polynomials to the twisted Baker-Akhiezer functions.
Recently, in a private communication Oleg Chalykh suggested that these twisted Baker-Akhiezer functions
could provide eigenfunctions of the commuting Hamiltonians associated with the $(-1,a)$ rays of the Ding-Iohara-Miki algebra.
In the paper, we discuss this suggestion and some evidence in its support.
\end{abstract}

\bigskip

\newcommand\smallpar[1]{
  \noindent $\bullet$ \textbf{#1}
}

\section{Introduction}

Theory of Macdonald polynomials \cite{Mac} remains one of the intriguing chapters of modern mathematical physics.
These polynomials have a number of remarkable properties, which however still escape clear group theory explanations.
On the other hand, they remain important players in integrable and superintegrable (in the sense of \cite{MMsi}) models of the Ruijsenaars type \cite{RS,Rui}. This makes the study of these special functions an always-inspiring field of research.

A fresh entity is here a set of (super)integrable systems \cite{MPSh,Max,China3,MMPdim} associated
with rays of commuting operators (Hamiltonians) within the quantum toroidal (or Ding-Iohara-Miki (DIM) $U_{q,t}(\widehat{\widehat{\mathfrak{gl}}}_1)$ algebra) \cite{DI,Miki1,AKMMMZ},
the rays being related with each other by Miki automorphisms \cite{Miki2,Miki1}.
A natural hope is that the set of their eigenfunctions is provided by Macdonald theory,
and indeed some results of this kind were already reported in \cite{MMPdim}.
Here we move this story a little further, being inspired by a letter from O. Chalykh
suggesting to look at the literature around the Cherednik-Macdonald-Mehta (CMM) formulas \cite{MM-conj,Ch1,EK}.
Of all approaches to these formulas, there is one \cite{ChE} most preferable for our purposes\footnote{This approach is basically a generalization of an earlier construction \cite{CV,Ves,Cha2} of eigenfunctions of the rational and trigonometric Calogero-Sutherland systems to the Ruijsenaars case.}. It originates
from the following observation.

Consider the Macdonald polynomial $M_R(\vec x;q,t)$, which is a
symmetric homogeneous polynomial in $N$ variables $x_i$, $i=1,\ldots,N$ labeled by the partition (Young diagram) $R$ with $N$ parts (lines), some of them may be zero. Since the Macdonald polynomial is homogeneous, it essentially depends on $N-1$ variables (up to a simple monomial factor). In particular,
for $N=2$ the polynomial $M_{[n_1,n_2]}(x_1,x_2;q,t)$ essentially depends only on the ratio $\frac{x_1}{x_2}$.
Looking at explicit expressions for these polynomials, e.g.

{\footnotesize
\be
M_{[8]}(x_1,x_2;q,t) =
x_1^8 + x_1^7x_2\frac{(q + 1)(q^2 + 1)(q^4 + 1)\boxed{(t - 1)}}{q^7t - 1}
+ x_1^6x_2^2\frac{(q^2 + 1)(q^6 + q^5 + q^4 + q^3 + q^2 + q + 1)(q^4 + 1)\boxed{(t - 1)(qt - 1)}}{(q^6t - 1)(q^7t - 1)}
+ \nn \\
+ x_1^5x_2^3\frac{(q + 1)(q^2 + 1)(q^2 - q + 1)(q^4 + 1)(q^6 + q^5 + q^4 + q^3 + q^2 + q + 1)
\boxed{(t - 1)(qt - 1)(q^2t - 1)}}{(q^7t - 1)(q^6t - 1)(q^5t - 1)}
+ \nn \\
+ x_1^4x_2^4\frac{(q^2 - q + 1)(q^4 + q^3 + q^2 + q + 1)(q^4 + 1)(q^6 + q^5 + q^4 + q^3 + q^2 + q + 1)
\boxed{(t - 1)(qt - 1)(q^2t - 1)(q^3t - 1)}}{(q^7t - 1)(q^6t - 1)(q^5t - 1)(q^4t - 1)}
+ \nn \\
+ x_1^3x_2^5\frac{(q + 1)(q^2 + 1)(q^2 - q + 1)(q^4 + 1)(q^6 + q^5 + q^4 + q^3 + q^2 + q + 1)\boxed{(t-1)(qt - 1)(q^2t - 1)}}
{(q^7t - 1)(q^6t - 1)(q^5t - 1)}
+\nn \\
+ x_1^2x_2^6\frac{(q^2 + 1)(q^6 + q^5 + q^4 + q^3 + q^2 + q + 1)(q^4 + 1)\boxed{(t - 1)(qt - 1)}}{(q^6t - 1)(q^7t - 1)}
+ x_1x_2^7\frac{(q + 1)(q^2 + 1)(q^4 + 1)\boxed{(t - 1)}}{q^7t - 1} + x_2^8
\nn
\ee
}

\noindent
it is clear that they are further simplified when $t=q^{-m}$ with $m<n/2$ (pay attention at the factors in boxes).
Then $(x_1x_2)^{-n\over 2}M_{[n]}(x_1,x_2;q,t)$ reduces to a sum of two degree-$m$ polynomials of the ratio $\xi:=\frac{x_1}{x_2}$: $\xi^{n/2-m}P_m(n,\xi)$ and $\xi^{m-n/2}P_m(n,\xi^{-1})$.
For example,
{\footnotesize
$$
P_2(8,\xi) = \xi^2 + \frac{(q + 1)(q^2 + 1)(q^4 + 1)\boxed{(q^{-2} - 1)}}{q^{5} - 1}\xi
+ \frac{(q^2 + 1)(q^6 + q^5 + q^4 + q^3 + q^2 + q + 1)(q^4 + 1)\boxed{(q^{-2} - 1)(q^{-1} - 1)}}{(q^4 - 1)(q^5 - 1)}
$$
}

\noindent
Despite these polynomials depend only on $q$, there is an entire family of them with different $m$,
which actually encodes the entire $t$ dependence.

These new polynomials nicknamed  (after \cite{CFV}) the Baker-Akhiezer (BA) functions in \cite{Cha}
possess a number of remarkable properties allowing to consider them as fundamental  building blocks
for the Macdonald polynomials. The most convenient way is to define these polynomials in shifted variables:
\be\label{shift}
\overline{\Psi}_m(\lambda,z)\sim P_m(n=\lambda+m,q^z)
\ee
with a proper normalization factor (see (\ref{psiM1}) below).
We also let $\lambda$ be an arbitrary (not obligatory integer) number. Then, these polynomials are symmetric:
\be
\boxed{
\overline{\Psi}_m(z,\lambda )=\overline{\Psi}_m(\lambda,z )
}
\label{sym}
\ee
despite the meaning of these variables is completely different.
The symmetry (\ref{sym}) can be considered as a precursor
for the mysterious Macdonald's duality relation (see \cite[sec.6,eq.(6.6)]{Mac})
\be
{M_\lambda(\{q^{\mu_i}t^{-i}\};q,t)\over M_\lambda(\{t^{-i}\};q,t)}={M_\mu(\{q^{\lambda_i}t^{-i}\};q,t)\over M_\mu(\{t^{-i}\};q,t)}
\label{duaid}
\ee
which extends them to non-integer values of parameters. In section 2, we explain that the Baker-Akhiezer function is generally a continuation of the Macdonald polynomials to $N$ non-integer complex parameters, and it is symmetric under permutation of these $N$ parameters with $N$ $z_i$, see formula (\ref{22}).

Perhaps, the most interesting is another set of identities,
\be
\boxed{
\overline{\Psi}_m(j,\lambda) = \overline{\Psi}_m(-j,\lambda), \ \ \  j=1,\ldots,m
}
\label{leq}
\ee
which allows one to {\it define} the polynomials
\be
\overline{\Psi}_m(z,\lambda) =q^{{1\over 2}\lambda z}\sum_{k=0}^mq^{{1\over 2}(m-2k)z}\psi_{m,\lambda,k}
\label{polPsi}
\ee
and then this set of identities (\ref{leq}) fixes these polynomials uniquely
up to an arbitrary function of $\lambda$, which is restored from the symmetry property (\ref{sym}).

Remarkable is the fact that the easy-to-solve linear relation (\ref{leq}) is nearly trivial,
and contains no references to the Macdonald polynomials.
Still it {\bf can be used to {\it define} them}\footnote{
For instance, at $m=1$ substitution of (\ref{polPsi}) into (\ref{leq}) gives (see (\ref{shift}))
$$\overline{\Psi}_1 = \psi_{1,\lambda,0}\xi^{\frac{\lambda+1}{2}} \left(
1- \frac{q^{\frac{\lambda+1}{2}}-q^{-\frac{\lambda+1}{2}}}{q^{\frac{\lambda-1}{2}}-q^{-\frac{\lambda-1}{2}}}\xi^{-1}
\right)
=  \psi_{1,\lambda,0}\xi^{\frac{\lambda+1}{2}} \left(
1- \frac{q^{\lambda+1}-1}{q^{ \lambda-1} -1}\cdot \frac{1}{q\xi}
\right)
$$
and the coefficient $-\frac{q^{\frac{\lambda+1}{2}}-q^{-\frac{\lambda+1}{2}}}{q^{\frac{\lambda-1}{2}}-q^{-\frac{\lambda-1}{2}}}$
at $t=q^{-1}$ and $\lambda=r-m=8-1=7$ coincides with $\left.\frac{(q + 1)(q^2 + 1)(q^4 + 1){(t - 1)}}{q^7t - 1}\right|_{t=q^{-1}}
\!\!\!= -\frac{q^4-q^{-4}}{q^3-q^{-3}}$,  clearly seen in the expansion of $M_{[8]}(x_1,x_2;q,t)$.
The remaining coefficient $ \psi_{1,\lambda,0} = q^{\frac{\lambda}{2}-1}-q^{-\frac{\lambda}{2}}$
is restored from the {\it additional} symmetry requirement (\ref{sym}) not imposed by the expansion of $M_{[8]}(x_1,x_2;q,t)$.
}, by analytical continuation in $t=q^{-m}$.
It is thus a natural desire to look for a realization of the mysteries of Macdonald theory
starting from the definition (\ref{leq}) of the BA function $\Psi_m(z,\lambda)$.

\bigskip

In this paper, we study an additional conjecture made by O. Chalykh \cite{Chapc}
that the deformations of the BA function called {\it twisted BA} \cite[Appendix by O. Chalykh]{ChE},
which were early used in generalizations of the CMM formulas \cite{Ch1},
can actually provide eigenfunctions for the commuting Hamiltonians associated with commutative DIM rays in the $N$-body (vector) representation \cite{MMPdim} (actually the relations \cite[Eqs.(5.6)-(5.12)]{ChF} along with \cite[Eq.(60)]{MMPdim} state exactly this for particular ray $(-1,2)$).

The paper is organized as follows. In section 2, we describe the construction of the BA functions starting from the simplest example of the symmetric function in two variables, $N=2$ and then extend it to generic $N$. In the case of $N=2$, a general simple expression for the BA function is available, while at arbitrary $N$, it is much more involved \cite[Eq.(3.16)]{Cha}. In order to get a flavour of what the BA function is in $N>2$ case, we describe its generic structure in the simplest non-trivial case of $N=3$ in section 3. In section 4, we introduce following \cite[Appendix by O. Chalykh]{ChE} the twisted BA function, and evaluate it in the simplest case at $N=2$. In fact, both the BA functions and the twisted BA functions are the main players in a set of integral identities that are called CMM formulas \cite{ChE}, and we discuss them in section 5. At last, section 6 contains the main point of the present paper: we test the conjecture that the twisted BA functions are proportional to eigenfunctions of the set of Hamiltonians from commutative subalgebras of the DIM algebra associated with $(-1,r)$ rays \cite{MMPdim}. Section 7 contains some concluding comments, and the Appendix illustrates the next non-trivial example of the twisted BA function at $N=2$.

\paragraph{Notation.} We use the standard definitions of the $q$-number:
$$
[n]_q:={q^n-1\over q-1}
$$
$q$-factorial:
$$
[n]_q!:=\prod_{i=1}^n[i]_q
$$
and $q$-binomial coefficient:
$$
\left[\begin{array}{c}  n \\ k \end{array} \right]_{q}:={[n]_q!\over [k]_q![n-k]_q!}
$$

We denote the Macdonald polynomial, which is a symmetric polynomial of the variables $x_i$, either through $M_R(\vec x;q,t)$ or through $M_R(\{x_i\};q,t)$. The Macdonald polynomial can also be treated as a graded polynomial of the power sums $p_k=\sum_{i=1}^Nx_i^k$. In this case, we use the notation $M_R\{p;q,t\}$.

\section{Macdonald-inspired family of BA functions}

\subsection{BA function at $N=2$}

The notion of BA function which are eigenfunctions of the Macdonald type Hamiltonians was introduced in \cite{Cha}. Let us first consider the $A_1$ case, i.e. that of two variables $x_1$ and $x_2$. Then, the Macdonald polynomial is labelled by the partition with two parts $[n_1,n_2]$, however, it effectively reduces to dependence on $n=n_1-n_2$, and has the form
\be\label{Mac}
M_{[n_1,n_2]}(x_1,x_2;q,t)&=&x_1^{n_1}x_2^{n_2}\cdot \sum_{k=0}^\infty\xi^{-k}
\prod_{j=1}^{k}{(q^{n-j+1}-1)(q^{j-1}t-1)\over(q^{j}-1)(q^{n-j}t-1)}=\nn\\
&=&(x_1x_2)^{n_1+n_2\over 2}\cdot \xi^{n\over 2}\sum_{k=0}^n\xi^{-k}
\prod_{j=1}^{k}{(q^{n-j+1}-1)(q^{j-1}t-1)\over(q^{j}-1)(q^{n-j}t-1)},\ \ \ \ \ \ \ \ \ \ \xi:={x_1\over
x_2}
\ee
The sum actually runs up to $k=n$, all other terms vanish because of the product vanishing.

In this section and sometimes in the next sections, we omit the trivial $U(1)$-factor $(x_1x_2)^{n_1+n_2/2}$ when dealing with the $N=2$ case in order to have formulas depending on the single variable $\xi$ (or $z$ below). In the case of generic $N$, we always keep this factor.

Now choose $t=q^{-m}$, and assume that $n$ is larger than $2m$. Then, of all coefficients in (\ref{Mac}), only those with $k=0,\ldots,m$ and with $k=n,n-1,\ldots,n-m$ are non-zero. Hence, (\ref{Mac}) is a sum of two terms: $\xi^{n/2-m}P_m(\xi)$ and $\xi^{m-n/2}P_m(\xi^{-1})$, where $P_m$ is a degree $m$ polynomial. Let us pick up one of these two terms,
\be
\xi^{n/2}\sum_{k=0}^m\xi^{-k}
\prod_{j=1}^{k}{(q^{n-j+1}-1)(q^{j-1}t-1)\over(q^{j}-1)(q^{n-j}t-1)}=
\xi^{n/2}\sum_{k=0}^m\xi^{-k}
\prod_{j=1}^{k}{(q^{n-j+1}-1)(q^{j-1-m}-1)\over(q^{j}-1)(q^{n-j-m}-1)}
\ee
choose the exponential parametrization of the variable: $\xi=q^{z}$, denote $\lambda=n-m$ and let $\lambda$ be an arbitrary complex number. Then, we define
\be\label{psi2}
\overline{\Psi}_m(z,\lambda)={\cal N}_\lambda\cdot q^{{1\over 2}\lambda z}\sum_{k=0}^mq^{{1\over 2}(m-2k)z}
\prod_{j=1}^{k}{(q^{\lambda+m-j+1}-1)(q^{j-1-m}-1)\over(q^{j}-1)(q^{\lambda-j}-1)}=
{\cal N}_\lambda\cdot q^{{1\over 2}\lambda z}\sum_{k=0}^mq^{{1\over 2}(m-2k)z}\psi_{m,\lambda,k}
\ee
where ${\cal N}$ is a normalization factor to be fixed later, and the bar over $\Psi$ refers to omitted $U(1)$-factor.
This quantity is called the BA function \cite{Cha}, and is nothing but the basis hypergeometric series \cite{Koorn}.

The Macdonald polynomial is given by the sum
\be
M_{[\lambda_1+m,\lambda_2+m]}(q^{z};q,q^{-m})\sim\overline{\Psi}_m(z,\lambda_1-\lambda_2)
+\overline{\Psi}_m(-z,\lambda_1-\lambda_2)
\ee
Let us discuss the basic properties of the BA function:
\begin{itemize}
\item One may note that
\be\label{symm}
\overline{\Psi}_m(z+j,\lambda)=\overline{\Psi}_m(z-j,\lambda)\ \ \ \ \ \hbox{for}\ j=1,2,\ldots,m\ \ \ \  \hbox{at}\  q^{z}=1
\ee
\item Note that the BA function is also an eigenfunction of the Ruijsenaars Hamiltonians. In particular,
\be\label{9}
\left({1-q^{z-m}\over 1-q^{z}}e^{\partial_z}+{1-q^{-z-m}\over 1-q^{-z}}e^{-\partial_z}\right)\overline{\Psi}_m(z,\lambda)=
q^{-{\lambda+m\over 2}}(q^{\lambda}+1)\overline{\Psi}_m(z,\lambda)
\ee
\item The properly normalized function $\Psi_m(z,\lambda)$ is also symmetric in variables $z$ and $\lambda$: 
\be
\overline{\Psi}_m(z,\lambda)=\overline{\Psi}_m(\lambda,z)
\ee
This is achieved by the choice of ${\cal N}$ to be
\be\label{Norm}
{\cal N}_\lambda=\prod_{j=1}^m\Big(q^{\lambda/2-j}-q^{-\lambda/2}\Big)
\ee
This property lies in the origin of the duality of the Macdonald polynomials (\ref{duaid}) \cite[sec.6,eq.(6.6)]{Mac}.
\end{itemize}

Finally, the relation with the Macdonald polynomial at integer $\lambda$ is
\be\label{psiM1}
M_{[\lambda+m]}(x_1,x_2;q,q^{-m})=(x_1x_2)^{(\lambda+m)/2}\cdot
\prod_{j=1}^m\Big(q^{{\lambda\over 2}-j}-q^{-{\lambda\over 2}}\Big)^{-1}
\cdot\left[\Big(\overline{\Psi}_m(z,\lambda)+\overline{\Psi}_m(-z,\lambda)\Big)\right]_{q^z={x_1\over x_2}}
\ee

As usual for the Macdonald related systems \cite{NS}, there is a Poincare symmetry $t\to q/t$, i.e. one can make all the construction steps for $t=q^{m+1}$. The corresponding Macdonald polynomial is given by the formula
\be\label{psiM2}
M_{[\lambda-m-1]}(x_1,x_2;q,q^{m+1})&=&(x_1x_2)^{(\lambda-m-1)\over 2}\cdot\prod_{j=1}^m\Big(q^{{\lambda\over 2}-j}-q^{-{\lambda\over 2}}\Big)^{-1}\times \nn\\
&\times&\prod_{j=-m}^m\left(q^{j}\sqrt{x_1\over x_2}-\sqrt{x_2\over x_1}\right)^{-1}
\cdot \left[\Big(\overline{\Psi}_m(z,\lambda)-\overline{\Psi}_m(-z,\lambda)\Big)\right]_{q^z={x_1\over x_2}}
\ee

Note that at $m=0$ these formulas reduce to the Weyl formulas:
\be
M_{[\lambda]}(x_1,x_2;q,t=1)=(x_1x_2)^{\lambda\over 2}\Big(q^{z\lambda\over 2}+q^{-{z\lambda\over 2}}\Big)=x_1^\lambda+x_2^\lambda=\mathfrak{m}_\lambda\\
M_{[\lambda]}(x_1,x_2;q,t=q)=(x_1x_2)^{\lambda\over 2}{q^{z(\lambda+1)\over 2}-q^{-{z(\lambda+1)\over 2}}\over q^{z\over 2}-q^{-{z\over 2}} }={x_1^{\lambda+1}-x_2^{\lambda+1}\over x_1-x_2}=S_\lambda(x_1,x_2)
\ee
where $\mathfrak{m}_R$ is the monomial symmetric polynomial, and $S_R$ is the Schur function, which is a character of the $A_{N-1}$ group (in this concrete case, $N=2$).

\subsection{BA function at arbitrary $N$}

Consider now the case of many complex variables $x_i$, $i=1,\ldots,N$ and $\lambda_i$, $i=1,\ldots,N$. We also use the notation
$\vec \lambda$, and the vector $\vec z$ with components
$z_j$, $j=1,\ldots,N$ parameterized as $q^{z_j}=x_j$. Then, we define the BA function as follows: it is a function
\be\label{gsBA}
\Psi_m(\vec z,\vec\lambda)=q^{(\vec\lambda+m\vec\rho) \vec z}\ \sum_{k_{ij}=0}^mq^{-\sum_{i>j}k_{ij}(z_i-z_j)}\psi_{m,\vec\lambda,k}
\ee
where $\vec\rho$ is the Weyl vector, i.e. $\vec\rho\cdot\vec z={1\over 2}\sum_{i=1}^N(N-2i+1)z_i$ and
$\Psi_m(x,\lambda)$ celebrates the property
\be\label{symm1}
\Psi_m(z_k+j,\vec\lambda)=\Psi_m(z_l+j,\vec\lambda)\ \ \ \ \  \forall k,l\ \ \hbox{and}\ \ 1\le j\le m\ \ \ \ \ \hbox{at}\ \ q^{z_k}=q^{z_l}
\ee
Surprisingly, these conditions are restrictive enough in order to fix the BA function uniquely up to a normalization. In particular, one can prove \cite{Cha} that
\begin{itemize}
\item such BA function is an eigenfunction of the Ruijsenaars Hamiltonian:
\be
\left(\sum_i\prod_{j\ne i}{q^{z_j}-q^{-m-z_i}\over q^{z_j}-q^{z_i}}e^{\p_{z_i}}\right)\Psi_m(\vec z,\vec \lambda)=q^{-{m\over 2}}\left(\sum_iq^{\lambda_i}\right)\Psi_m(\vec z,\vec \lambda)
\ee
\item Upon a proper normalization, which is given by the coefficient at the leading term equal to
\be
\psi_{m,\vec\lambda,0}=\prod_{k>l}\prod_{j=1}^m\Big(q^{{\lambda_k-\lambda_l\over 2}-j}-q^{-{\lambda_k-\lambda_l\over 2}}\Big)
\ee
the BA function is symmetric:
\be\label{22}
\Psi_m(\vec z,\vec\lambda)=\Psi_m(\vec\lambda,\vec z)
\ee
\item Evidently, it has the properties:
\be
\Psi_m(w\vec z,w\vec\lambda)&=&\Psi_m(\vec z,\vec\lambda),\ \ \ \ \ w\in W \ \ \text{(equivariance)}\\
\Psi_m(-\vec z,-\vec\lambda)&=&\Psi_m(\vec z,\vec\lambda) \ \ \ \ \ \ \ \ \ \ \ \ \ \ \ \ \ \
\text{(reflection)}\\
\Psi_m(-\vec z,\vec\lambda)&=&\Psi_m(\vec z,\vec\lambda)\Big|_{q\to q^{-1}}
\ \ \ \ \ \ \ \ \ \text{(involution)}
\ee
where $W$ is the Weyl group of $A_{N-1}$.
\end{itemize}

Note that, in the case of $N=2$, this quantity $\Psi_m$ (\ref{gsBA}) and the quantity $\overline{\Psi}_m$ (\ref{psi2}) are related by the formula
$$
\boxed{
\Psi_m(z_1,z_2;\lambda_1,\lambda_2)=q^{(z_1+z_2)(\lambda_1+\lambda_2)\over 2}\ \overline{\Psi}_m(z_1-z_2,\lambda_1-\lambda_2)}
$$

One can obtain Macdonald polynomials by taking the sums (at integer $\lambda_i$
ordered as $\lambda_1 \geq \dots \geq \lambda_N$)
\be
M_{\vec\lambda+m\vec\rho}(\{x_i\};q,q^{-m})=
\prod_{k>l}\prod_{j=1}^m\Big(q^{{\lambda_k-\lambda_l\over 2}-j}-q^{-{\lambda_k-\lambda_l\over 2}}\Big)^{-1}\cdot
\left[\sum_{w\in W}\Psi_m(w\vec z,\vec\lambda)\right]_{q^{z_j}=x_j}\\
M_{\vec\lambda-(m+1)\vec\rho}(\{x_i\};q,q^{m+1})=
\prod_{k>l}\prod_{j=1}^m\Big(q^{{\lambda_k-\lambda_l\over 2}-j}-q^{-{\lambda_k-\lambda_l\over 2}}\Big)^{-1}\cdot
\prod_{k>l}\prod_{j=-m}^m\left(q^{j}\sqrt{x_k\over x_l}-\sqrt{x_l\over x_k}\right)^{-1}\times\nn\\
\times\left[\sum_{w\in W}(-1)^w\Psi_m(w\vec z,\vec\lambda)\right]_{q^{z_j}=x_j}
\ee
Since the Macdonald polynomial is the Weyl-invariant sum of the BA functions, and the Ruijsenaars Hamiltonian is also Weyl invariant, it is not surprising that the BA function is the eigenfunction of this Hamiltonian along with the Macdonald polynomial. The essential difference is, however, that the BA function provides us with the eigenfunction with arbitrary eigenvalues, while the Macdonald polynomials are eigenfunctions with eigenvalues parameterized by the integer partitions only.

\section{Generic structure
of the BA function at $N\ne 2$}

In contrast with the case of $N=2$, at generic $N$ the BA function is given by more involved formulas (see, e.g., \cite[Eq.(3.16)]{Cha}). Here we discuss the general structure of the BA function at the simplest non-trivial example of $N=3$.
For the purposes of this section, recall that $x_i = q^{z_i}$ and further introduce $\Lambda_i = q^{\lambda_i}$.

One may note that the formula for $\overline{\Psi}_m(z,\lambda)$ \eqref{psi2}
in the $N=2$ case is quite explicit and manifest. Moreover,
modulo overall $q^{\lambda z}$ factor, dependence on $q^\lambda$ of coefficients
in front of the \textit{modes} $q^{k z}$ is factorized into simple $q^\lambda$-linear
Pochhammer-like factors: immediately suggesting generalization to arbitrary, non-integral $t=q^{-m}$.

Still, at $N=3$ this structure is, at first glance, lost. Here note that the ansatz
\eqref{gsBA} is generally overdetermined: some monomials in $q^z_i, i=1..N$
get a sum of $\psi_{m,\vec\lambda,k}$, not just one. In other words, in contrast with the $N=2$ case, when the number of equations (\ref{symm}) just matches the number of unknown coefficients (up to a common normalization), at $N>2$ the number of equations (\ref{symm1})  is not enough in order to fix the coefficients $\psi_{m,\vec\lambda,k}$ in \eqref{gsBA} uniquely. However, it turns out that {\it all} possible solutions to these equations gives rise to one and the same BA function (again up to a common normalization), and one can choose a solution most conveniently parameterizing the coefficients $\psi_{m,\vec\lambda,k}$.

To be more concrete, at $N=3$ we have the following set of positive roots:
\begin{align} \label{eq:n-3-pos-roots}
  \underbrace{r_1 = e_1 - e_2, r_2 = e_2 - e_3}_{\text{simple roots}},\ \ r_3 = e_1 - e_3,
\end{align}
where, clearly, $r_3 = r_1 + r_2$ is a non-simple root. The sum in \eqref{gsBA} runs over $k_{12}$, $k_{23}$ and $k_{13}$, and we use the notation $\psi_{m,\vec\lambda,k_{12},k_{23},k_{13}}$ for the coefficients of \eqref{gsBA}. Then, the coefficient
of monomial $q^{z_1 - z_3}$ is proportional to
\begin{align} \label{eq:sum-two-psis}
  \psi_{1,\vec\lambda,1,1,0} + \psi_{1,\vec\lambda,0,0,1},
\end{align}
and there is no way to distinguish these two $\psi$'s based on periodicity equations
\eqref{symm1}. Moreover, their sum turns out to be equal to
\begin{align}
  \eqref{eq:sum-two-psis} =
  {\scriptscriptstyle
    \frac{(q-1)(q^4 \Lambda_1 \Lambda_2 \Lambda_3 + 2 q^3 \Lambda_1 \Lambda_2 \Lambda_3 - q^2 \Lambda_1^2 \Lambda_2
      -q^2 \Lambda_1^2 \Lambda_3-q^2 \Lambda_1 \Lambda_2^2-q^2 \Lambda_1 \Lambda_3^2-q^2 \Lambda_2^2 \Lambda_3
      -q^2 \Lambda_2 \Lambda_3^2 + 2 q \Lambda_1 \Lambda_2 \Lambda_3 + \Lambda_1 \Lambda_2 \Lambda_3)}{(q \Lambda_3-\Lambda_2)(q \Lambda_2-\Lambda_1) q (q \Lambda_3-\Lambda_1)}
  }
\end{align}
i.e., at first glance, to nothing good in particular.

\bigskip

One may depict the structure of $\Psi$ coefficients for various values
of $m$ on the Newton's plane:
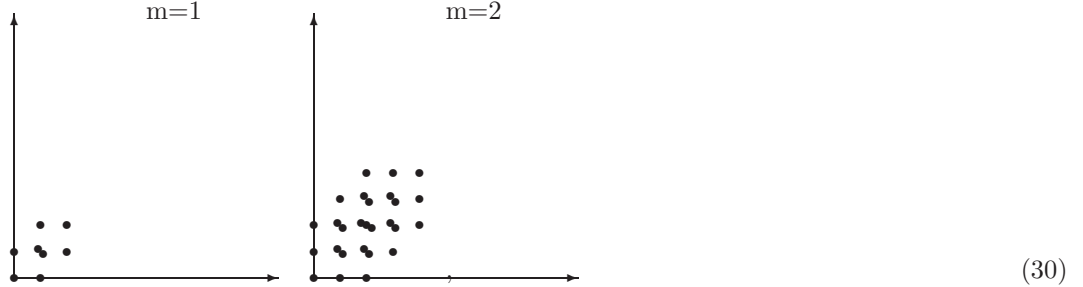
\begin{align} \label{eq:psi-newton-polytopes}
  \begin{picture}(100,100)
    \put(0,98){m=1}
    \put(-50,0){\vector(0,1){100}}
    \put(-50,0){\vector(1,0){100}}
    \put(-50,0){\put(0,0){\circle*{3}}
      \put(10,0){\circle*{3}}
      \put(0,10){\circle*{3}}
      \put(20,10){\circle*{3}}
      \put(10,20){\circle*{3}}
      \put(20,20){\circle*{3}}
      \put(10,10){
        \put(-1,1){\circle*{3}}
        \put(1,-1){\circle*{3}}
      }
    }
  \end{picture}
  \ \ \ \
  \begin{picture}(100,100)
    \put(0,98){m=2}
    \put(-50,0){\vector(0,1){100}}
    \put(-50,0){\vector(1,0){100}}
    \put(-50,0){
      \put(0,0){\circle*{3}}
      \put(10,0){\circle*{3}}
      \put(20,0){\circle*{3}}
      \put(0,10){\circle*{3}}
      \put(10,10){
        \put(-1,1){\circle*{3}}
        \put(1,-1){\circle*{3}}
      }
      \put(20,10){
        \put(-1,1){\circle*{3}}
        \put(1,-1){\circle*{3}}
      }
      \put(30,10){\circle*{3}}
      \put(0,20){\circle*{3}}
      \put(10,20){
        \put(-1,1){\circle*{3}}
        \put(1,-1){\circle*{3}}
      }
      \put(20,20){
        \put(-2,1){\circle*{3}}
        \put(0,0){\circle*{3}}
        \put(2,-1){\circle*{3}}
      }
      \put(30,20){
        \put(-1,1){\circle*{3}}
        \put(1,-1){\circle*{3}}
      }
      \put(40,20){\circle*{3}}
      \put(10,30){\circle*{3}}
      \put(20,30){
        \put(-1,1){\circle*{3}}
        \put(1,-1){\circle*{3}}
      }
      \put(30,30){
        \put(-1,1){\circle*{3}}
        \put(1,-1){\circle*{3}}
      }
      \put(40,30){\circle*{3}}
      \put(20,40){\circle*{3}}
      \put(30,40){\circle*{3}}
      \put(40,40){\circle*{3}}
    },
  \end{picture}
\end{align}
where the number of dots at a given node denotes number of different $\psi-$coefficients in front of the same $x_i-$monomial.
Clearly there are such monomials that, nevertheless, have just one $\psi$-factor:
the ones on the boundary of the Newton polytope \eqref{eq:psi-newton-polytopes}.
It turns out that the corresponding $\psi$'s \textit{are} factorized and manifestly equal to
\begin{align}\label{eq:psi-singlet-n-3}
  \psi_{m,\vec{\lambda},a,b,c} \Bigg{|}_{\text{singlet}} = & \
  \left[
    \begin{array}{c}
      m \\ a
    \end{array}
    \right]_{1/q}
  \frac{\prod_{j=1}^a
    \left(q^{m-j+1} \Lambda_1 - \Lambda_2\right)}
       {\prod_{j=1}^a\left(
         q^{j} \Lambda_2 - \Lambda_1
         \right)}
  \left[
    \begin{array}{c}
      m \\ b
    \end{array}
    \right]_{1/q}
  \frac{\prod_{j=1}^b
    \left(q^{m-j+1} \Lambda_1 - \Lambda_3\right)}
       {\prod_{j=1}^b\left(
         q^{j} \Lambda_3 - \Lambda_1
         \right)}
       \\ \notag
       & \ \left[
    \begin{array}{c}
      m \\ c
    \end{array}
    \right]_{1/q}
  \frac{\prod_{j=1}^c
    \left(q^{m-j+1} \Lambda_2 - \Lambda_3\right)}
       {\prod_{j=1}^c\left(
         q^{j} \Lambda_3 - \Lambda_2
         \right)}
\end{align}
and this ``singlet'' expression readily generalizes to arbitrary $N$
\begin{align}\label{eq:psi-singlet-arbitrary-n}
  \psi_{m,\vec{\lambda},k} \Bigg{|}_{\text{singlet}} = & \
  \prod_{i < j}
  \left[
    \begin{array}{c}
      m \\ k_{i,j}
    \end{array}
    \right]_{1/q}
  \frac{\prod_{l=1}^{k_{i,j}}
    \left(q^{m-l+1} \Lambda_i - \Lambda_j\right)}
       {\prod_{l=1}^{k_{i,j}}\left(
         q^{l} \Lambda_j - \Lambda_i
         \right)}
\end{align}
where the $q$-binomial coefficient is
made from $q$-numbers with inverse $q$.

\bigskip

\bigskip

The way to restore the factorizability structure in \eqref{eq:sum-two-psis}
(and in general for $N=3$) is to extract the singlet factor from $\psi$'s
\begin{align}
  \psi_{1,\vec\lambda,1,1,0} =\psi_{1,\vec\lambda,1,1,0} \Bigg{|}_{\text{singlet}}
  \cdot \widetilde{\psi}_{1,\vec\lambda,1,1,0}
  \\ \notag
  \psi_{1,\vec\lambda,0,0,1} = \psi_{1,\vec\lambda,0,0,1} \Bigg{|}_{\text{singlet}}
  \cdot \widetilde{\psi}_{1,\vec\lambda,0,0,1}
\end{align}
and to demand that $\widetilde{\psi}_{1,\vec\lambda,\vec k}$
depends \textit{only} on non-simple root $\Lambda_1 - \Lambda_3$.
The solution is then
\begin{align}
  \widetilde{\psi}_{1,\vec\lambda,1,1,0}=
  \frac{1}{q}
  \frac{(q^2 \Lambda_3-\Lambda_1) (q \Lambda_1-\Lambda_3)}{(\Lambda_1-\Lambda_3)(q \Lambda_3-\Lambda_1)}
  \ \ \ \ \ \ \ \ \ \ \ \ \
  \widetilde{\psi}_{1,\vec\lambda,0,0,1} =
  \frac{(-1)}{q}
  \frac{(q^2 \Lambda_1-\Lambda_3)(q \Lambda_3-\Lambda_1)}
       {(q \Lambda_1-\Lambda_3)(\Lambda_1-\Lambda_3)}
\end{align}

\bigskip

For arbitrary $m$, the first few answers are
\begin{align}
  \widetilde{\psi}_{m,\vec\lambda,0,0,1} = & \
  \frac{(-1)}{q}
  \frac{(q^{2 m} \Lambda_1 - \Lambda_3)}
       {(q^m \Lambda_1 - \Lambda_3)(q^{m-1} \Lambda_1 - \Lambda_3)}
       \cdot (q \Lambda_3 - \Lambda_1)
       \\ \notag
       \widetilde{\psi}_{m,\vec\lambda,1,1,0} = & \
       \frac{1}{q}
       \frac{(q^{m} \Lambda_1 - \Lambda_3)}
            {(q^{m-1} \Lambda_1 - \Lambda_3)}
            \cdot \frac{(q^2 \Lambda_3 - \Lambda_1)}{(q \Lambda_3 - \Lambda_1)}
            \\ \notag
            \widetilde{\psi}_{m,\vec\lambda,0,1,1} =
            \widetilde{\psi}_{m,\vec\lambda,1,0,1} = & \
            \frac{(-1)}{q}
            \frac{(q^{2 m-1} \Lambda_1 - \Lambda_3)}
                 {(q^m \Lambda_1 - \Lambda_3)(q^{m-2} \Lambda_1 - \Lambda_3)}
                 \cdot (q \Lambda_3 - \Lambda_1)
                 \\ \notag
            \widetilde{\psi}_{m,\vec\lambda,1,2,0} =
            \widetilde{\psi}_{m,\vec\lambda,2,1,0} = & \
       \frac{1}{q}
       \frac{(q^{m} \Lambda_1 - \Lambda_3)}
            {(q^{m-2} \Lambda_1 - \Lambda_3)}
            \cdot \frac{(q^3 \Lambda_3 - \Lambda_1)}{(q \Lambda_3 - \Lambda_1)}
\end{align}

\bigskip

\def\lbb{\left[\!\left[ }
    \def\rbb{\right]\!\right] }

With some care, one can deduce a general formula for the extra factor
$\widetilde{\psi}_{m,\vec\lambda,a,b,c}$ with arbitrary multiplicities $a,b,c$
\begin{align}
  \widetilde{\psi}_{m,\vec\lambda,a,b,c} =
  \frac{\lbb2 m-a-b\rbb}{\lbb2 m -a-b-c\rbb}
  \cdot
  \frac{\lbb m-1-a-b-c\rbb}{\lbb m -1-a-b\rbb}
  \cdot
  \frac{\lbb m-a-b\rbb}{\lbb -1-a-b\rbb}
  \cdot
  \frac{\lbb -1-a\rbb}{\lbb m-a\rbb}
  \cdot
  \frac{\lbb -1-b\rbb}{\lbb m-b\rbb}
  \cdot
  \frac{\lbb m-c\rbb}{\lbb m-1\rbb},
\end{align}
where $\lbb n \rbb$ is the infinite downward Pochhammer-like product
\begin{align}
  \lbb n \rbb = \prod_{i=0}^\infty \left(q^{n - i} \Lambda_1 - \Lambda_3\right)
  =: \left(q^n \Lambda_1;\Lambda_3\right)_{-\infty}
  =: \left(q^n\right)^{-}_{1,3}
\end{align}

\bigskip

\newcommand\pochlzm[1]{\left(#1\right)_{1,3}^-}

Therefore, it is also possible to define the $\psi$-function at arbitrary $t = q^{-m}$
by
\begin{align}
  \widetilde{\psi}_{t,\vec\lambda,a,b,c} =
  \frac{\pochlzm{t^{-2} q^{-a-b}}}{\pochlzm{t^{-2}q^{-a-b-c}}}
  \cdot
  \frac{\pochlzm{t^{-1} q^{-1-a-b-c}}}{\pochlzm{t^{-1} q^{-1-a-b}}}
  \cdot
  \frac{\pochlzm{t^{-1}q^{-a-b}}}{\pochlzm{q^{-1-a-b}}}
  \cdot
  \frac{\pochlzm{q^{-1-a}}}{\pochlzm{t^{-1}q^{-a}}}
  \cdot
  \frac{\pochlzm{q^{-1-b}}}{\pochlzm{t^{-1}q^{-b}}}
  \cdot
  \frac{\pochlzm{t^{-1} q^{-c}}}{\pochlzm{t^{-1} q^{-1}}}
\end{align}

\begin{align}\label{eq:psi-singlet-n-3-arb-t}
  \psi_{t,\vec{\lambda},a,b,c} \Bigg{|}_{\text{singlet}} = & \
  \prod_{s\in(a,b,c)} t^{s} q^{-s}
  \cdot
  \frac{\left(t^{-1};1\right)_{-\infty}\left(q^{-1-s};1\right)_{-\infty}}
       {\left(q^{-1};1\right)_{-\infty}
         \left(t^{-1}q^{-s};1\right)_{-\infty}}
  \cdot \frac{\left(t^{-1}\right)_{s_1,s_2}^-}{\left(t^{-1}q^{-s}\right)_{s_1,s_2}^-}
  \cdot \frac{\left(q^{-1-s}\right)_{s_1,s_2}^-}{\left(q^{-1}\right)_{s_1,s_2}^-},
\end{align}
where $s_1$ and $s_2$ denote the indices of $\Lambda$ of the corresponding root,
i.e. if $s=a$ then $s_1 = 1$, $s_2 = 2$.

\section{Twisted BA function\label{tBA}}

Twisted BA function is defined as a sum
\be\label{BAtN}
\Psi_m^{(a)}(\vec z,\vec\lambda)=q^{{\vec\lambda\cdot\vec z\over a}+m\vec\rho\cdot \vec z}\ \sum_{k_{ij}=0}^{ma}q^{-\sum_{i>j}{k_{ij}\over a}(z_i-z_j)}\psi^{(a)}_{m,\vec\lambda,k}
\ee
with the property
\be
\Psi_m^{(a)}(z_k+j,\vec\lambda)=\varepsilon^j\Psi_m^{(a)}(z_l+j,\vec\lambda)\ \ \ \ \  \forall k,l\ \ \hbox{and}\ \ 1\le j\le m\ \ \ \ \ \hbox{at}\ \ \varepsilon q^{z_k\over a}= q^{z_l\over a}
\ee
for any $\varepsilon$ such that $\varepsilon^a=1$.

This BA function is still defined unambiguously up to a normalization, and is symmetric, $\Psi_m^{(a)}( z,\lambda)=\Psi_m^{(a)}(\lambda, z)$ with a proper normalization. It is also a common eigenfunction of integrable Weyl-invariant Hamiltonians.

\paragraph{An example of $N=2$.} One can explicitly solve the conditions for the BA function in the case of $N=2$,
\be
\overline{\Psi}_m^{(a)}(z+j,\lambda)=\varepsilon^j\overline{\Psi}_m^{(a)}(z-j,\lambda)\ \ \ \ \  \forall \ 1\le j\le m\ \ \ \ \ \hbox{at}\ \ \varepsilon q^{z\over a}=1
\ee
and obtain for
\be\label{BAt2}
\overline{\Psi}_m^{(a)}(z,\lambda)={\cal N}_\lambda^{(a)}\cdot q^{{1\over 2a}(\lambda +ma)\cdot z}\ \sum_{k=0}^{am}q^{- {kz\over a}}\psi^{(a)}_{m,\lambda, k}
\ee
at $a=1$
\be
\psi^{(1)}_{m,\lambda, k}=(-1)^kq^{-mk+{k(k-1)\over 2}}{[m]_q!\over[m-k]_q![k]_q!}\prod_{i=1}^k{[\lambda+m-i+1]_q\over
[\lambda-i]_q}
\ee
at $a=2$
\be\label{ba2c}
\psi^{(2)}_{m,\lambda, k}=\sum_{r=max(0,k-m)}^{[k/2]}(-1)^{k+r}q^{({k\over 2}-r)(\lambda-2m-1)-mr+{r(r-1)\over 2}}{[m+k-2r]_q!\over[k-2r]_q![m-k+r]_q!}
{\prod_{i=1}^r[\lambda+m-i+1]_q\over[r]_q!\prod_{i=1}^{k-r}[\lambda-i]_q}
\ee
where $[x]$ denotes the integer part of $x$,
and $[-n]_q!$ at positive $n$ is understood as having a pole, hence, automatic restricting $r\ge k-m$, and the restrictions on the summing range over $r$ in (\ref{ba2c}) can be lifted. At integer $\lambda$, (\ref{ba2c}) can be rewritten in terms of the $q$-binomial coefficients:
$$
\psi^{(2)}_{m,\lambda, k}=\sum_{r}(-1)^{k+r}q^{({k\over 2}-r)(\lambda-2m-1)-mr+{r(r-1)\over 2}}\cdot
{\left[\begin{array}{c}  \lambda+m \\ r \end{array} \right]_{q}\cdot\left[\begin{array}{c}  m+k-2r \\ k-2r \end{array} \right]_{q}\over\left[\begin{array}{c}  \lambda-1 \\ r-k \end{array} \right]_{q}\cdot\left[\begin{array}{c}  m-k+r \\ \lambda-1 \end{array} \right]_{q}}
$$
At non-integer $\lambda$, this formula is realized by the $q$-$\Gamma$-functions.

Now let us note that the symmetry of $\overline{\Psi}_m^{(a)}(z,\lambda)$ w.r.t. the permutation of $z$ and $\lambda$ is provided by choosing in (\ref{BAt2}) the same normalization factor as in (\ref{Norm}):
\be
{\cal N}_\lambda^{(a)}={\cal N}_{\lambda}
\ee

One can definitely construct a twisted counterpart of the Macdonald polynomial. To this end, one removes the normalization factor and continues the twisted BA functions from $t=q^{-m}$ to arbitrary $t$:
\be\label{Mt}
\overline{\Psi}^{(a)}(z,\lambda;q,t)\longrightarrow q^{{1\over 2a}\lambda\cdot z}t^{-{z\over 2}}\ \sum_{k=0}q^{- {kz\over a}}\psi^{(a)}_{\lambda, k}(q,t)
\ee
where, for instance, at $a=2$, one has
\be
\psi^{(2)}_{\lambda, k}(q,t)=\sum_{r=0}(-t)^{r-k}{q^{({k\over 2}-r)(\lambda-1)+{r(r-1)\over 2}}\over[k-2r]_q![r]_q!}\prod_{i=r-k}^{k-2r}
{tq^i-1\over q-1}
{\prod_{i=1}^r(q^{\lambda-i+1}t^{-1}-1)\over\prod_{i=1}^{k-r}[\lambda-i]_q}
\ee
Now, one makes the replace $q^\lambda\to q^nt^a$ with integer $n$ which makes the sum (\ref{BAt2}) finite with the cut-off at $k\ge n$, and considers the sum over action of the Weyl group. At the same example of $a=2$:
\be
M^{(2)}_{[n]}(x_1,x_2;q,t)&\sim&\overline{\Psi}^{(2)}(z,\lambda;q,t)+\overline{\Psi}^{(2)}(-z,\lambda;q,t)\sim\nn\\
&\sim&\sum_{k=0}\left(q^{({n\over 2}-k){z\over 2}}+
q^{(k-{n\over 2}){z\over 2}}\right)\times\nn\\
&\times&t^{-{k\over 2}}\cdot\sum_{r=0}(-1)^{r-k}{q^{({k\over 2}-r)(n-1)+{r(r-1)\over
2}}\over[k-2r]_q![r]_q!}\prod_{i=r-k}^{k-2r}{tq^i-1\over q-1}
{\prod_{i=1}^r(q^{n-i+1}-1)\over\prod_{i=1}^{k-r}(q^{n-i}t-1)}
\ee
which, upon making the substitution $q^{z\over 2}={x_1\over x_2}$ and multiplying with the factor $(x_1x_2)^{n\over 2}$ similarly to (\ref{psiM1}), becomes a symmetric polynomial of two variables $x_1$ and $x_2$. The same procedure definitely works at arbitrary $a$ and $N$, however, the structure of the answer already at $N=2$, $a=3$ becomes much more involved, see the Appendix. More examples are available at \cite{MMPf}.

\section{Integral formulas}

\paragraph{Two-particle case.} Let us calculate the integral of product of the two-particle BA functions (\ref{psi2}):
\be
I(\lambda,\mu)=\int_{-\infty}^\infty dz q^{-{z^2\over 4}}{\overline{\Psi}_m(z,\lambda)\overline{\Psi}_m(z,\mu)\over
\prod_{j=1}^m(1-q^{j+z})(1-q^{j-z})}
\ee
The pole terms in the integrand are understood as a geometric series.

Another way to calculate this integral is to reduce it to the contour integral with an additional $\theta$-function as above:
\be
I(\lambda,\mu)=\oint{d\xi\over\xi}{\overline{\Psi}_m(\log\xi,\lambda)\overline{\Psi}_m(\log\xi,\mu)\over
\prod_{j=1}^m(1-q^j\xi)(1-q^j\xi^{-1})}\sum_{k\in\mathbb{Z}}q^{k^2}\xi^k
\ee
This integral is proportional to $\overline{\Psi}_m(\lambda,\mu)$:
\be
I(\lambda,\mu)=(-1)^mq^{-m(m+1)}q^{\lambda^2+\mu^2\over 4}\overline{\Psi}_m(\lambda,\mu)
\ee

\paragraph{$N$-particle case.} This integral is immediately generalized to the generic case:
\be
\int_{-\infty}^\infty \ldots\int_{-\infty}^\infty \left(\prod_{i=1}^{N}dz_i q^{-{z_i^2\over 2}}\right)
{\Psi_m(\vec z,\vec\lambda)\Psi_m(\vec z,\vec\mu)\over\prod_{k\ne l}\prod_{j=1}^m(1-q^{j+z_k-z_l})}=\nn\\
=\oint\ldots\oint\prod_{i=1}^{N}\left(
{d\xi_i\over\xi_i}\sum_{k\in\mathbb{Z}}q^{k^2\over 2}\xi_i^k\right)
{\Psi_m(\log\xi_i,\vec\lambda)\Psi_m(\log\xi_i,\vec\mu)\over
\prod_{k\ne l}\prod_{j=1}^m(1-q^{j+z_k-z_l})}=\nn\\
=(-1)^mq^{-m(m+1)}q^{|\lambda|^2+|\mu|^2\over 2}\Psi_m(\vec\lambda,\vec\mu)
\ee

\paragraph{Twisted BA function case.}
These formulas can be further generalized to produce the twisted BA function in the two-particle case:
\be
\int_{-\infty}^\infty dz q^{-{az^2\over 4}}{\overline{\Psi}_m(z,\lambda)\overline{\Psi}_m(z,\mu)\over
\prod_{j=1}^m(1-q^{j+z})(1-q^{j-z})}&=&\oint{d\xi\over\xi}{\overline{\Psi}_m(\log\xi,\lambda)\overline{\Psi}_m(\log\xi,\mu)\over
\prod_{j=1}^m(1-q^j\xi)(1-q^j\xi^{-1})}\sum_{k\in\mathbb{Z}}q^{k^2/a}\xi^k=\nn\\
&=&(-1)^mq^{-m(m+1)}q^{\lambda^2+\mu^2\over 4a}\overline{\Psi}_m^{(a)}(\lambda,\mu)
\ee
and for generic $N$:
\be
\int_{-\infty}^\infty \ldots\int_{-\infty}^\infty \left(\prod_{i=1}^{N}dz_i q^{-{az_i^2\over 2}}\right)
{\Psi_m(\vec z,\vec\lambda)\Psi_m(\vec z,\vec\mu)\over\prod_{k\ne l}\prod_{j=1}^m(1-q^{j+z_k-z_l})}=\nn\\
=\oint\ldots\oint\prod_{i=1}^{N}\left(
{d\xi_i\over\xi_i}\sum_{k\in\mathbb{Z}}q^{k^2\over 2a}\xi_i^k\right)
{\Psi_m(\log\xi_i,\vec\lambda)\Psi_m(\log\xi_i,\vec\mu)\over
\prod_{k\ne l}\prod_{j=1}^m(1-q^{j+z_k-z_l})}=\nn\\
=(-1)^mq^{-m(m+1)}q^{|\lambda|^2+|\mu|^2\over 2a}\Psi_m^{(a)}(\vec\lambda,\vec\mu)
\ee
These integrals are basically an origin of the CMM formulas \cite{MM-conj,Ch1} and were first realized in \cite{ChE}.

\section{BA functions as eigenfunctions}

\subsection{On the eigenfunctions of DIM Hamiltonians}

Let us briefly discuss the issue of eigenfunctions of the DIM Hamiltonians in the Fock and $N$-body representations following \cite{MMPdim}.

In the Fock representation, the representation space is spanned by polynomials, since it is given by action of $p_k$ and ${\p\over \p p_k}$. On the other hand, in the $N$-body representation, this is no longer the case, since the action of operators of multiplication with $x_i^{-1}$ gets the polynomials out of this space. Surprisingly enough, the vertical ray Hamiltonians still have polynomial eigenfunctions (the Macdonald polynomials).
However, the eigenfunctions of Hamiltonians associated with ray (-1,1) are already very different in the Fock and $N$-body representations: in the Fock representation, they are given by a kind of Itzykson-Zuber integral
\be
\Phi\{p,\bar p\}=\sum_\lambda q^{-{1\over 2}\nu_{\lambda^\vee}}t^{{1\over 2}\nu_\lambda}M_\lambda\{p;q,t\}\overline{M}_{\lambda^\vee}\{\bar p;q,t\}
\ee
where $\nu_\lambda:=2\sum_i(i-1)\lambda_i$, and $\lambda^\vee$ denotes the conjugate Young diagram,
while, in the $N$-body representation, they are given by the Macdonald polynomials multiplied with a very non-polynomial factor:
\be
\Phi(\vec x,R)=\left(\prod_{i=1}^Nx_i^{-{1\over 2}(\log_q x_i-1)}\right)M_R(\vec x;q,t)
\ee
Note that the first non-trivial Hamiltonian of this ray is simple in the $N$-body representation:
\be
H_1^{(x)}=\sum_i\prod_{j\ne i}{x_j-tx_i\over x_j-x_i}{1\over x_i}q^{x_i\partial_i}
\ee
and is more involved in the Fock one:
\be
H_1^{(p)}=\oint_0dz\exp\left(\sum_k{1-t^{-k}\over k}z^kp_k\right)\exp\left(-\sum_k(1-q^k)z^{-k}{\p\over\p p_k}\right)
\ee
Note that this situation differs from the $W_{1+\infty}$ algebra and from the Yangian algebra cases, where the eigenfunctions of the commutative rays in the Fock and $N$-body representations are related in a simple way \cite{MMMP1}.

\subsection{BA functions as eigenfunctions of DIM Hamiltonians}

Here we explain that the twisted BA functions are eigenfunctions of the DIM Hamiltonians for rays (-1,a) in the $N$-body representation of the DIM algebra. In \cite{MMPdim}, we discussed in detail how to construct Hamiltonians for commutative families in the $N$-body representation. In particular, the first Hamiltonians of the commutative series associated with rays (-1,a) are produced by repeated commutators with the Macdonald-Ruijsenaars Hamiltonian $\hat H_{MR}$ \cite{Mac,Rui}
\be\label{rc}
\hat H_1^{(-1,a+1)}=[\hat H_1^{(-1,a)},\hat H_{MR}]
\ee
where the initial condition Hamiltonian is
\be
H_1^{(-1,0)}=\sum_{i=1}^Nx_i^{-1}
\ee
and
\be
\hat H_{MR}={\sqrt{q}\over q-1}\ \sum_{i=1}^N\prod_{j\ne i}{tx_i-x_j\over x_i-x_j}q^{\hat D_i}
\ee
with
\be
    \hat D_i:&=&x_i{\p\over\p x_i}
\ee
Note that we rescaled here all the Hamiltonians with a factor of ${\sqrt{q}\over q-1}$ as compared with \cite{MMPdim}.

The first Hamiltonian of the DIM commutative family associated with ray (-1,2) \cite{MMPdim} is manifestly \cite[eqs.(5.10)-(5.12)]{ChF}\footnote{The authors of this paper considered also a more general Hamiltonian \cite[eqs.(5.18)]{ChF}, which corresponds to the cone commutative family $\beta e_{(-1,0)}+\alpha e_{(-1,1)}+e_{(-1,2)}$ in \cite[sec.4.5]{MMPdim}, i.e. to an arbitrary linear combination of rays  (-1,0),  (-1,1) and (-1,2).},\cite[eq.(60)]{MMPdim}
\be\label{12N}
\hat H^{(-1,2)}_1&= & \ \sum_{i=1}^N \frac{1}{q^{1\over 2}x_i}
    \prod_{j \neq i} \frac{(t x_i - x_j)}{(x_i - x_j)}\frac{(qt x_i - x_j)}{(qx_i - x_j)}
    \ttop{i}\ttop{i}
    \nn \\
    & + &{q^{1\over 2}(t - q)(t - 1)\over q-1}
    \sum_{i \neq j} \prod_{k \neq i,j}
    \frac{(t x_i - x_k)(t x_j - x_k)}{(x_i - x_k)(x_j - x_k)}
    \frac{1}{(q x_i - x_j)} \ttop{i} \ttop{j}
\ee
As was proposed in \cite{ChF}, the eigenfunctions of this Hamiltonian (and of the whole ray) at $t=q^{-m}$ are proportional to the twisted BA functions $\Psi_m^{(2)}(\vec z,\vec\lambda)$ with $x_i=q^{z_i}$.

In particular, in the two-particle case, the Hamiltonian is
\be\label{12}
\hat H^{(-1,2)}_1(x_1,x_2)&= & \ \frac{1}{q^{1\over 2}x_1}
    \frac{(q^{-m} x_1 - x_2)}{(x_1 - x_2)}\frac{(q^{-m+1} x_1 - x_2)}{(qx_1 - x_2)}
    \ttop{1}\ttop{1}+ \ \frac{1}{q^{1\over 2}x_2}
    \frac{(q^{-m} x_2 - x_1)}{(x_2 - x_1)}\frac{(q^{-m+1} x_2 - x_1)}{(qx_2 - x_1)}
    \ttop{2}\ttop{2}
    \nn \\
    & + &{q^{1\over 2}(q^{-m} - q)(q^{-m} - 1)}{x_1+x_2\over (qx_1-x_2)(qx_2-x_1)}
\ttop{1} \ttop{2}
\ee
or
\be
\hat {\overline {H}}^{(-1,2)}_1(z)&= &\ {q^{-z}\over q}
    \frac{(q^{z-m} - 1)}{(q^z - 1)}\frac{(q^{z+1-m} - 1)}{(q^{z+1} - 1)}
    e^{2\p_z}+ \ {q^{z}\over q}
    \frac{(q^{-z-m} - 1)}{(q^{-z} - 1)}\frac{(q^{1-z-m} - 1)}{(q^{1-z} - 1)}
    e^{-2\p_z}
    \nn \\
    & + &{q^{1\over 2}(q^{-m} - q)(q^{-m} - 1)}{q^{z\over 2}+q^{-{z\over 2}}\over (q^{z+1}-1)(q^{1-z}-1)}
\ee
and one can check from the explicit formulas (\ref{BAt2}), (\ref{ba2c}) that
\be
\hat H^{(-1,2)}_1(q^{z_1},q^{z_2})\left[q^{{1\over 4}\sum_{i=1,2}z_i^2}\cdot\Psi_m^{(2)}(z_1-z_2,\lambda_1-\lambda_2)\right]=
q^{-m}\left(q^{\lambda_1}+q^{\lambda_2}\right)\left[q^{{1\over 4}\sum_{i=1,2}z_i^2}
\cdot\Psi_m^{(2)}(z_1-z_2,\lambda_1-\lambda_2)\right]\nn
\ee
and
\be
\hat {\overline H}_1^{(-1,2)}(z)\left[q^{{1\over 4}z^2}\cdot\overline{\Psi}_m^{(2)}(z,\lambda)\right]=q^{-m}\left(q^{\lambda\over 2}+q^{-{\lambda\over 2}}\right) \left[q^{{1\over 4}z^2}\cdot\overline{\Psi}_m^{(2)}(z,\lambda)\right]
\ee

The same scheme works for an arbitrary $N$: the eigenfunctions of the $N$-particle Hamiltonian (\ref{12N}) are proportional to the $N$-particle BA function,
\be
\hat H_1^{(-1,2)}(x_i=q^{z_i})\left[q^{{1\over 4}\sum_i z_i^2}\cdot\Psi_m^{(2)}(\vec z,\vec \lambda)\right]=q^{-m}
\left(\sum_iq^{\lambda_i}\right) \left[q^{{1\over 4}\sum_iz_i^2}\cdot\Psi_m^{(2)}(\vec z,\vec\lambda)\right]
\ee

Note that, in accordance with \cite[sec.6.2.2]{MMPdim}, at $a=1$, the eigenfunction of ray (-1,1) is given by the untwisted BA function\footnote{There is an additional factor of $q^{z\over 2}$ in this eigenfunction in \cite[sec.6.2.2]{MMPdim}, however, it just results in a simple additional factor in the eigenvalue. For the sake of simplicity, we do not include it here.}
\be
\hat H_1^{(-1,1)}(x_i=q^{z_i})\left[q^{{1\over 2}\sum_iz_i^2}\cdot\Psi_m(\vec z,\vec \lambda)\right]=q^{-{m\over 2}}\left(\sum_iq^{\lambda_i}\right) \left[q^{{1\over 2}\sum_iz_i^2}\cdot\Psi_m(\vec z,\vec \lambda)\right]
\ee

Similarly to these two formulas, one may expect that the twisted BA function $\Psi_m^{(a)}(\vec z,\vec \lambda)$ provides an eigenfunction for the Hamiltonians of ray $(-1,a)$ \cite[sec.6]{MMPdim}\footnote{The factor $q^{-{amk\over 2}}$ is consistent with the normalization of Hamiltonians in \cite{MMPdim}.}:
\be
\boxed{
\hat H^{(-1,a)}_k(x_i=q^{z_i})\left[q^{{1\over 2a}\sum_iz_i^2}\cdot\Psi_m^{(a)}(z,\lambda)\right]=
q^{-{amk\over 2}}\left(\sum_iq^{k \lambda_i}\right)\left[ q^{{1\over 2a}\sum_iz_i^2}\cdot\Psi_m^{(a)}(z,\lambda)\right]
}
\ee
We have verified this claim for a few first values of $N$, $a$ and $k$ with the computer.

All these formulas for the eigenfunctions and Hamiltonians are immediately continued from $t=q^{-m}$ to arbitrary $t$. Moreover, the twisted Macdonald polynomials constructed as in sec.\ref{tBA} are eigenfunctions of these Hamiltonians.

There is an important subtle issue here: when dealing with (twisted) Macdonald polynomials, one can obtain the (-1,a) ray Hamiltonians using explicit formulas like \cite[Eq.(52)]{MMPdim} expressing them through the Cherednik operator, or, equivalently, one can generate them by repeated commutators of the elements \cite[Eqs.(42)-(43)]{MMPdim}, since they coincide on the space of symmetric functions. However, when dealing with the BA functions, which are not symmetric functions, there is a difference between the two, and one has to use the ``true" DIM elements, which are obtained from the repeated commutators, i.e. to use the second way of doing. Another possibility is
to explicitly symmetrize the operators obtained with help of the Cherednik operators: the symmetrized version coincides with the DIM elements. For instance, in the notation of \cite{MMPdim}, the symmetrization looks as
\newcommand\br[1]{
  \left(#1\right)
}
\begin{align}
  \hat H^{(-1,a)}_k = \left(\sum_i \br{\widetilde{C}_i x_i \widetilde{C}_i}^k \right)_{\text{symm}}
  = \sum_{\sigma \in S_N} \sum_i \br{\sigma\br{\widetilde{C}_i} \sigma\br{x_i}
  \sigma\br{\widetilde{C}_i}}^k
\end{align}

\section{Conclusion}

To summarize, in this paper we reconsidered the six items of the Chalykh's approach to Macdonald polynomials:

\begin{itemize}
\item{} Splitting/decomposition of Macdonald polynomials $M_{\vec\lambda+m\vec\rho}(\vec x;q,t)$ at special points $t=q^{-m}$ and $t=q^{m+1}$ into a set
of conjugate $\lambda$-dependent polynomials of degree $m$ called ``BA functions" and defined at arbitrary (not obligatory integer) set of $\lambda$.

\item{} An alternative definition of the BA functions, ordinary and twisted, through a system of linear equations.

\item{} Representation of under-defined system expansion coefficients in a factorized form.

\item{} An alternative system of finite-difference equations for these BA functions,
identifying them as the eigenfunctions of commuting families of Hamiltonians,
which, in turn, are the families associated with commuting subalgebras of DIM
called ``rays" in \cite{MMPdim}.

\item{} Alternative systems of integral non-linear equations for the BA functions,
where the weights are quasiperiodic theta-functions defined on the infinite covering of torus,
which, in appropriate coordinates, have essential singularities at ``infinity".
Then, the integrals can be represented by an infinite sum of residues, which has a well-defined
meaning at $|q|<1$.

\item{} When the BA functions are substituted in the integral formulas by the Macdonald polynomials at $t=q^{n+1}$,
only a finite number of residues contributes to the integral relations so that they are easy to treat and check.
These relations are also known as the CMM formulas \cite{Ch1}, which are thus elegantly proved within the Chalykh's approach \cite{ChE}.

\end{itemize}

Altogether this constitutes an interesting way to re-define the Macdonald polynomials
as an analytical continuation in $t$ at integer points in $\lambda$ of a nice system of functions.
Moreover, these functions are exactly what is needed to establish relations with the DIM sets (rays) of
integrable systems.

In this paper, we just illustrated these points in an elementary way with a set of simple examples.An extension of this presentation to full generality, which preserves its conceptual and technical simplicity,
is still to be done.
Another direction are various generalizations, from the Macdonald polynomials to the elliptic Macdonald polynomials \cite{MMZ}
and further to the elliptic Shiraishi functions \cite{ELS}
or from hyperpolynomials of the Hopf link \cite{MMPhopf} to other knots and links.
As to the opposite direction of considering simpler cases like the Jack polynomials, see earlier papers \cite{Ves,Cha2}.

\section*{Acknowledgements}

We are grateful to Oleg Chalykh for attracting our attention to his wonderful results. The work was partially funded within the state assignment of the Institute for Information Transmission Problems of RAS. Our work is partly supported by grant RFBR 21-51-46010 ST-a, by the grants of the Foundation for the Advancement of Theoretical Physics and Mathematics ``BASIS".

\section*{Appendix}

Let us illustrate the structure of $\psi^{(a)}_{m,\lambda,k}$ at $N=2$, $a=3$. Expressions for this quantity at the first few values of $k$ are rather simple:
\be
\psi^{(3)}_{m,\lambda,1}&=&-q^{2\lambda-3m-2\over 3}{[m+1]_q!\over  [m-1]_q![\lambda-1]_q}\nn\\
\psi^{(3)}_{m,\lambda,2}&=&-q^{\lambda-3m-2\over 3}{[m+1]_q!\over  [m-1]_q![\lambda-1]_q}+q^{4\lambda-6m-5\over 3}
{[m+2]_q!\over [2]_q![m-2]_q![\lambda-1]_q[\lambda-2]_q}\nn\\
\psi^{(3)}_{m,\lambda,3}&=&-q^{-m}{[m]_q![\lambda+m]_q\over[m-1]_q![\lambda-1]_q}
+q^{\lambda-2m-2}{[m+2]_q!\over[m-2]_q![\lambda-1]_q[\lambda-2]_q}
-q^{2\lambda-3m-3}{[m+3]_q!\over[m-3]_q![3]_q![\lambda-1]_q[\lambda-2]_q[\lambda-3]_q}\nn\\
\psi^{(3)}_{m,\lambda,4}&=&q^{2\lambda-6m-2\over 3}{[m+1]_q!\over[m-2]_q!}{[\lambda+m]_q\over[\lambda-1]_q[\lambda-2]_q}+
q^{2\lambda-6m-5\over 3}{[m+2]_q!\over[m-2]_q![2]_q!}{1\over[\lambda-1]_q[\lambda-2]_q}-\nn\\
&-&q^{5\lambda-9m-11\over 3}{[m+3]_q!\over[m-3]_q![2]_q!}{1\over[\lambda-1]_q[\lambda-2]_q[\lambda-3]_q}+
q^{8\lambda-12m-14\over 3}{[m+4]_q!\over[m-4]_q![4]_q!}{1\over[\lambda-1]_q[\lambda-2]_q[\lambda-3]_q[\lambda-4]_q}\nn\\
\psi^{(3)}_{m,\lambda,5}&=&q^{\lambda-6m-2\over 3}{[m+1]_q!\over[m-2]_q!}{[\lambda+m]_q\over[\lambda-1]_q[\lambda-2]_q}-
q^{4\lambda-6m+4\over 3}{[m+2]_q!\over[m-3]_q![2]_q!}{1\over[\lambda-1]_q[\lambda-2]_q}-\nn\\
&-&q^{4\lambda-9m-11\over 3}{[4]_q\over [2]_q}{[m+3]_q!\over[m-3]_q![2]_q!}{1\over[\lambda-1]_q[\lambda-2]_q[\lambda-3]_q}+
q^{7\lambda-12m-17\over 3}{[m+4]_q!\over[m-4]_q![3]_q!}{1\over[\lambda-1]_q[\lambda-2]_q[\lambda-3]_q[\lambda-4]_q}-\nn\\
&-&q^{10\lambda-15m-20\over 3}{[m+5]_q!\over[m-5]_q![5]_q!}{1\over[\lambda-1]_q[\lambda-2]_q[\lambda-3]_q[\lambda-4]_q[\lambda-5]_q}\nn
\ee
We remind that ${1\over [-n]_q!}$ is defined to be zero at natural $n$. Hence, for instance, at $m=1$, only the first term in $\psi^{(3)}_{1,\lambda,1}$, $\psi^{(3)}_{1,\lambda,2}$ and $\psi^{(3)}_{1,\lambda,3}$ contributes, and all $\psi^{(3)}_{1,\lambda,k}$'s with $k>3$ are zero.

However, though the general structure of these formulas holds intact at higher $k$, some numerical coefficients already at $k=6$ become quite involved:
\be
\psi^{(3)}_{m,\lambda,6}&=&q^{-2m+1}{[m]_q!\over[m-2]_q!}{[\lambda+m]_q[\lambda+m-1]_q\over[2]_q![\lambda-1]_q[\lambda-2]_q}+\nn\\
&+&a_mq^{\lambda-4m+1}{[m+2]_q!\over[3]_q![m-3]_q!}{[\lambda+m]_q\over[\lambda-1]_q[\lambda-2]_q[\lambda-3]_q}
+q^{\lambda-4m+1}{[m+3]_q!\over[m-3]_q!}{[m-4]_q\over[\lambda-1]_q[\lambda-2]_q[\lambda-3]_q}+\nn\\
&+& q^{2\lambda-4m-6}{[4]_q\over [2]_q}{[m+4]_q!\over[m-4]_q![3]_q!}{1\over[\lambda-1]_q[\lambda-2]_q[\lambda-3]_q
[\lambda-4]_q}-\nn\\
&-&q^{3\lambda-5m-8}{[m+5]_q!\over[m-5]_q![4]_q!}{1\over[\lambda-1]_q[\lambda-2]_q[\lambda-3]_q
[\lambda-4]_q[\lambda-5]_q}+\nn\\
&+&q^{4\lambda-6m-9}{[m+6]_q!\over[m-6]_q![6]_q!}{1\over[\lambda-1]_q[\lambda-2]_q[\lambda-3]_q[\lambda-4]_q[\lambda-5]_q
[\lambda-6]_q}
\ee
and the coefficients in the second line of this formula are rather peculiar, $a_m$ being $m$-dependent linear combinations of $q$-numbers: $\boxed{a_m=q^{m+3}[m-3]_q-[m+1]_q-q^{m-2}[2]_q}$.

\end{document}